\documentclass[prl,twocolumn]{revtex4-2}
\usepackage{amsmath}
\usepackage{amssymb}
\usepackage{graphicx}
\usepackage{braket}
\usepackage{stmaryrd}
\usepackage{hyperref}
\usepackage{color}
\usepackage{dsfont}

\hypersetup{
    colorlinks,
    citecolor=blue,
    linkcolor=blue,
    urlcolor=blue
}

\begin{document}
\title{Thermodynamic Concentration Inequalities and Tradeoff Relations}
\author{Yoshihiko Hasegawa}
\email{hasegawa@biom.t.u-tokyo.ac.jp}
\affiliation{Department of Information and Communication Engineering, Graduate
School of Information Science and Technology, The University of Tokyo,
Tokyo 113-8656, Japan}

\author{Tomohiro Nishiyama}
\email{htam0ybboh@gmail.com}
\affiliation{Independent Researcher, Tokyo 206-0003, Japan}

\date{\today}
\begin{abstract}

Thermodynamic tradeoff relations quantify the fundamental concept of ``no free lunch'' in the physical world, suggesting that faster and more precise physical processes come at a higher thermodynamic cost.
The key elements in these tradeoff relations are the thermodynamic uncertainty relation and speed limit, which are closely tied to information inequalities from which other tradeoff relations are derived. 
Concentration inequalities are relations that complement information inequalities in statistical analyses and have been widely used in various fields.
However, their role in thermodynamic tradeoff relations remains unclear.
This Letter develops thermodynamic concentration inequalities that provide bounds for the distribution of observables in quantum and classical Markov processes. 
We derive a set of tradeoff relations that generalize speed limits and thermodynamic uncertainty relations from the developed thermodynamic concentration inequalities.
The derived tradeoff relations hold under minimal assumptions of the underlying physical processes. 
This Letter clarifies the role of concentration inequalities in thermodynamics, paving the way for deriving new tradeoff relations.

\end{abstract}
\maketitle

\textit{Introduction.---}The concept of tradeoff relations, illustrating the principle of no effortless gains in the physical world, has attracted considerable interest in thermodynamics.
The thermodynamic uncertainty relation \cite{Barato:2015:UncRel,Gingrich:2016:TUP,Garrahan:2017:TUR,Dechant:2018:TUR,Terlizzi:2019:KUR,Hasegawa:2019:CRI,Hasegawa:2019:FTUR,Dechant:2020:FRIPNAS,Vo:2020:TURCSLPRE,Koyuk:2020:TUR,Erker:2017:QClockTUR,Brandner:2018:Transport,Carollo:2019:QuantumLDP,Liu:2019:QTUR,Guarnieri:2019:QTURPRR,Saryal:2019:TUR,Hasegawa:2020:QTURPRL,Hasegawa:2020:TUROQS,Sacchi:2021:BosonicTUR,Kalaee:2021:QTURPRE,Vu:2021:QTURPRL}
(see \cite{Horowitz:2019:TURReview} for a review) and speed limit \cite{Mandelstam:1945:QSL,Margolus:1998:QSL,Deffner:2010:GenClausius,Taddei:2013:QSL,DelCampo:2013:OpenQSL,Deffner:2013:DrivenQSL,Pires:2016:GQSL,OConnor:2021:ActionSL,Shiraishi:2018:SpeedLimit,Ito:2018:InfoGeo,Ito:2020:TimeTURPRX,Vu:2021:GeomBound} (see \cite{Deffner:2017:QSLReview} for a review) are essential components of the tradeoff relations. The thermodynamic uncertainty relation implies that a thermodynamic machine requires a larger thermodynamic input to achieve a higher level of precision. The speed limit suggests that making more significant or faster changes to the state requires more thermodynamic or quantum resources.

Information inequalities, such as the Cram\'er-Rao inequality, are significant in statistical theory and have been widely utilized in the study of thermodynamic uncertainty relations \cite{Hasegawa:2019:CRI,Dechant:2019:MTUR,Hasegawa:2020:QTURPRL} and speed limits \cite{Frowis:2012:FisherQSL,Taddei:2013:QSL,Ito:2020:TimeTURPRX}. 
The Cram\'er-Rao inequality provides the lower bound of a statistical estimator $\hat{\vartheta}$, which is given by
\begin{align}
    \mathrm{Var}[\hat{\vartheta}]\ge\frac{1}{\mathcal{I}(\vartheta)},
    \label{eq:CramerRao_def}
\end{align}
where $\vartheta$ is the parameter to be estimated and $\mathcal{I}(\vartheta)$ is the Fisher information. 
Equation~\eqref{eq:CramerRao_def} states that the precision of the estimator $\hat{\vartheta}$ is fundamentally limited regardless of how the estimator is constructed.
Concentration inequalities constitute another pivotal class of statistical tools that, similar to information inequalities, underpin the theoretical foundation of statistical analysis 
\cite{Boucheron:2013:ConcIneqBook,Zhang:2021:ConcIneqReview}.
The most well-known instance of concentration inequality is the Markov inequality. 
Let $X$ be a random variable and let $\mathbb{E}[X]$ be the expectation of $X$. 
For a positive constant $a$, the Markov inequality states
\begin{align}
    P(|X|\geq a)\leq\frac{\mathbb{E}\left[|X|\right]}{a}.
    \label{eq:Markov_inequality}
\end{align}
The Markov inequality is a key concept in probability theory and statistics because of its generality; it does not
necessitate knowledge of the probability distribution except for the
mean. 
However, despite the generality and prevalence of concentration inequalities in numerous areas, including machine learning, statistics, and optimization, they have received little attention in studying thermodynamic tradeoff relations.

\begin{figure}
\centering
\includegraphics[width=1.0\linewidth]{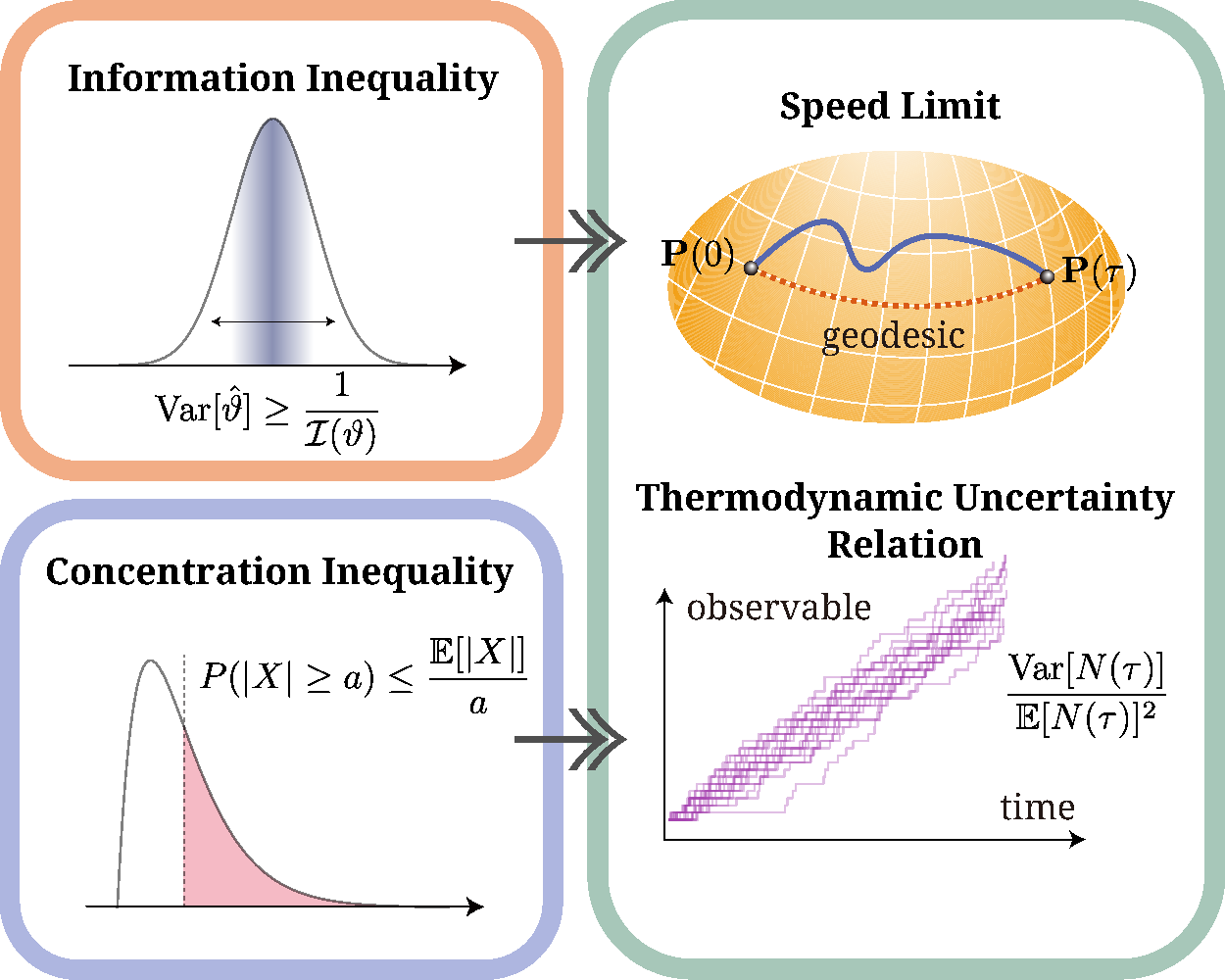}
\caption{
Illustrative connection between inequalities and tradeoff relations. 
Information inequalities, such as the Cram\'er-Rao inequality, are instrumental in deriving speed limits and thermodynamic uncertainty relations. 
This Letter establishes concentration inequalities as another powerful tool for establishing thermodynamic tradeoff relations.
}
\label{fig:concept}
\end{figure}

This Letter derives the thermodynamic concentration inequalities that provide lower bounds for the probability distribution of observables in quantum and classical Markov processes.
This lower bound comprises the dynamical activity, 
central in thermodynamic tradeoff relations \cite{Garrahan:2017:TUR,Shiraishi:2018:SpeedLimit,Terlizzi:2019:KUR,Hasegawa:2020:QTURPRL,Hasegawa:2023:BulkBoundaryBoundNC}. 
From the thermodynamic concentration inequalities, we establish a family of relations that provide bounds for classical and quantum Markov processes. The derived bounds lead to generalization of the speed limits and thermodynamic uncertainty relations (Fig.~\ref{fig:concept}). 
The primary distinction of this study from previous research lies in the application of concentration inequalities. This approach allow us to derive trade-off relations that were not previously established, which are formulated using the $p$-norm.
The derivation of the thermodynamic concentration inequality opens the way to discovering new tradeoff relations in classical and quantum thermodynamics.

\textit{Methods.---}The concentration inequalities for thermodynamics can be considered in open quantum dynamics governed by the Lindblad equation, which describes the evolution of a quantum system coupled to an external environment.
Let $\rho_S(t)$ be the density operator at time $t$. 
The Lindblad equation is represented as $\dot{\rho}_S=\mathcal{L} \rho_S$, where $\mathcal{L}$ is the Lindblad superoperator:
\begin{align}
    \mathcal{L} \rho_S\equiv -i[H_S, \rho_S]+\sum_{m=1}^{N_C} \mathcal{D}[L_m] \rho_S,
    \label{eq:superop_def}
\end{align}
where $H_S$ denotes the system Hamiltonian, $L_m$ denotes the $m$th jump operator, $N_C$ denotes the number of jump channels, and $\mathcal{D}[L] \rho_S\equiv $ $L \rho_S L^{\dagger}-\frac{1}{2}\{L^{\dagger} L, \rho_S\}$. 
The jump operators describe various dissipative processes such as energy dissipation, dephasing, and decoherence.

The Lindblad equation in Eq.~\eqref{eq:superop_def} is also capable of describing classical Markov processes. 
Specifically, consider a classical Markov process comprising $M$ states, denoted by a set $\mathfrak{B} \equiv \{B_1,B_2,\cdots,B_M\}$. 
Let $P(\nu;t)$ be the probability that the state equals $B_\nu$ at time $t$ and
let $W_{\nu\mu}$ be the transition rate from state $B_\mu$ to state $B_\nu$. 
The dynamics of the classical Markov process is described by
\begin{align}
\frac{d}{dt}\mathbf{P}(t)=\mathbf{W}\mathbf{P}(t),
    \label{eq:Master_eq_def}
\end{align}
where $\mathbf{P}(t) \equiv [P(1;t),\cdots,P(M;t)]^\top$ and the diagonal elements of $\mathbf{W} \equiv [W_{\nu\mu}]$ are defined by $W_{\mu\mu}=-\sum_{\nu\neq\mu}W_{\nu\mu}$. 
Then by setting $\rho_S(t) = \mathrm{diag}[P(1;t),\ldots,P(M;t)]$, $H_S=0$, and $L_{\nu\mu} = \sqrt{W_{\nu\mu}}\ket{B_\nu}\bra{B_\mu}$, 
where $\ket{B_\mu}$ denotes the orthogonal basis corresponding to the classical state $B_\mu$,
the Lindblad equation of Eq.~\eqref{eq:superop_def} represents the classical dynamics of Eq.~\eqref{eq:Master_eq_def}.

\begin{figure}
\includegraphics[width=1\linewidth]{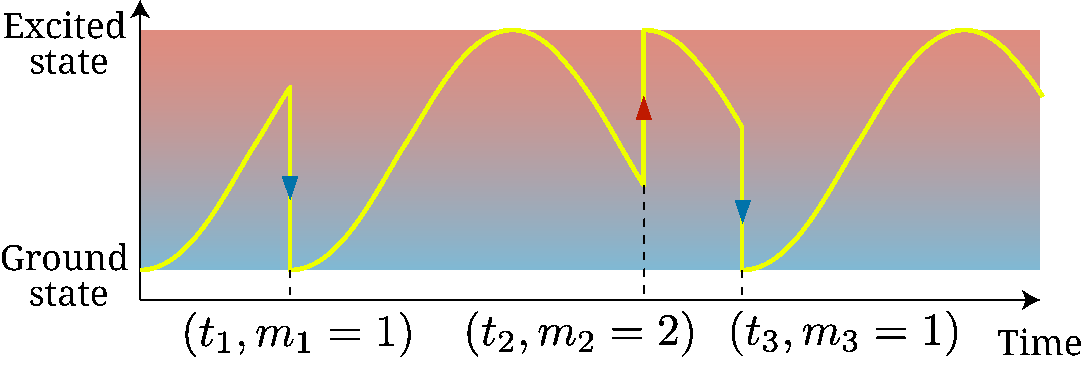}
\caption{
Continuous measurement and trajectory. 
Two-level system comprising the excited state $\ket{\mathfrak{e}}$ and ground state $\ket{\mathfrak{g}}$. 
The Hamiltonian is $H_S \propto \ket{\mathfrak{e}}\bra{\mathfrak{g}} + \ket{\mathfrak{g}}\bra{\mathfrak{e}}$ and two jump operators are $L_1 \propto \ket{\mathfrak{g}}\bra{\mathfrak{e}}$ and $L_2 \propto \ket{\mathfrak{e}}\bra{\mathfrak{g}}$. 
During the continuous measurement, 
the Hamiltonian $H_S$ generates smooth dynamics, while the jump operators $L_1$ and $L_2$ cause transitions to the ground and excited states, respectively. 
The occurrence of jump types $m=1$, $2$, and $1$ are at $t=t_1$, $t_2$, and $t_3$ in that order. 
The observable $N_m(\tau)$, which counts the number of $m$th jumps within $[0,\tau]$, is $N_1(\tau)=2$ and $N_2(\tau)=1$. 
}
\label{fig:quantum_traj_explanation}
\end{figure}

Consider continuous measurement in the Lindblad equation corresponding to continuous monitoring of the environment coupled to the system (see Ref.~\cite{Supp:2024:ConcenIneq}). 
\nocite{Landi:2023:CurFlucReviewPRXQ,Verstraete:2010:cMPS,Osborne:2010:Holography,Uhlmann:1992:BuresGeodesic}
The measurement record of the continuous measurement
consists of the type of jump event and its time stamp. 
Suppose that the measurement record has $K$ jump events within the interval $[0,\tau]$. 
where $m_k \in \{1,\cdots,N_C\}$ represents the jump type at time $t_k$. 
Moreover, it is possible to measure the system after the continuous measurement. 
Let $s$ be the output of the measurement applied to the system at $t=\tau$. 
Then, the measurement record can be expressed as
\begin{align}
    \zeta_\tau \equiv \left[\left(t_1, m_1\right), \left(t_2, m_2\right), \ldots, \left(t_K, m_K\right);s\right].
    \label{eq:traj_zeta_def}
\end{align}
Here, the sequence $\zeta_\tau$ is referred to as a trajectory (Fig.~\ref{fig:quantum_traj_explanation}). 
Given the trajectory, the system dynamics can be described using a stochastic process. 
In classical Markov processes, $\zeta_\tau$ as defined in Eq.~\eqref{eq:traj_zeta_def} completely determines the stochastic time evolution. Consequently, the trajectory can be represented by
\begin{align}
    \zeta_\tau = \{X(t) | 0 \le t \le \tau\},
    \label{eq:traj_classical_def}
\end{align}
where
$X(t) \in \mathfrak{B}$ is the state at time $t$
(see Ref.~\cite{Supp:2024:ConcenIneq} for details).  

Next, we define the observables for the continuous measurement. 
Let $N(\zeta_\tau)$ be an observable of the trajectory $\zeta_\tau$, which is the key quantity in this Letter. 
The observable $N(\zeta_\tau)$ can be arbitrary as long as the following condition is satisfied:
\begin{align}
    N(\zeta_{\varnothing}) = 0,
    \label{eq:Ovarnothing_def}
\end{align}
where $\zeta_{\varnothing}$ denotes the trajectory with no jumps. 
In the following, we simplify the notation $N(\zeta_\tau)$ to $N(\tau)$.
An example of $N(\tau)$ that satisfies Eq.~\eqref{eq:Ovarnothing_def} is the counting observable:
\begin{align}
    N(\tau)=\sum_{m}C_{m}N_{m}(\tau),
    \label{eq:M_def}
\end{align}
where $[C_{m}]$ is the real weight vector and $N_m(\tau)$ is the number of $m$th jumps within $[0,\tau]$. 
For example, the thermodynamic uncertainty relation in Ref.~\cite{Garrahan:2017:TUR} concerns the counting observable in Eq.~\eqref{eq:M_def}. 
Since $\zeta_\tau$ is a random variable, $N(\tau)$ is also a random variable.
Therefore, we can consider its probability distribution $P(N(\tau))$, which plays a central role in this Letter.

Here, we introduce the relevant thermodynamic quantities. 
We first review classical dynamical activity, a central cost term in thermodynamic tradeoff relations \cite{Garrahan:2017:TUR,Shiraishi:2018:SpeedLimit,Terlizzi:2019:KUR}. 
The classical dynamical activity was originally proposed as an order parameter in the field of glass transitions \cite{Biroli:2013:GlassTransition}. 
In classical Markov processes, the time-integrated dynamical activity within $[0, \tau]$ is defined as $\mathcal{A}_{\mathrm{cl}}(\tau)\equiv\int_{0}^{\tau}\mathfrak{a}(t)dt$, where $\mathfrak{a}(t)$ is the dynamical activity at $t$:
\begin{align}
    \mathfrak{a}(t)\equiv\sum_{\nu,\mu\,(\nu\ne\mu)}P(\mu;t)W_{\nu\mu}.
    \label{eq:mathfrak_a_def}
\end{align}
$\mathcal{A}_\mathrm{cl}(t)$ evaluates the average number of jump events within $[0, \tau]$ and thus it quantifies the activity of the system. 
Using the jump operator $L_m$, classical dynamical activity can be represented by
\begin{align}
    \mathcal{A}(\tau) \equiv \int_{0}^{\tau}\sum_{m}\mathrm{Tr}_{S}[L_{m}\rho_{S}(t)L_{m}^{\dagger}]dt,
    \label{eq:classical_DA_def2}
\end{align}
where $\mathrm{Tr}_S$ is the trace with respect to the system. 
The equivalence between $\mathcal{A}_\mathrm{cl}(\tau)$ and $\mathcal{A}(\tau)$ for the classical limit is confirmed by assigning $L_m = L_{\nu\mu} = \sqrt{W_{\nu \mu}}\ket{B_\nu}\bra{B_\mu}$ in Eq.~\eqref{eq:classical_DA_def2}, and considering the density operator $\rho_S(t)$ whose diagonal entries follow the probability distribution $P(\mu;t)$.

The concept of dynamical activity can be generalized to the open quantum dynamics described by the Lindblad equation, which plays a central role in tradeoff relations \cite{Hasegawa:2020:QTURPRL,Hasegawa:2023:BulkBoundaryBoundNC,
Nakajima:2023:SLD,Nishiyama:2024:ExactQDAPRE,Nishiyama:2024:OpenQuantumRURJPA}. 
Let $\mathcal{B}(\tau)$ be the quantum generalization of $\mathcal{A}(\tau)$ in Eq.~\eqref{eq:classical_DA_def2}, called the quantum dynamical activity. 
Because the classical dynamical activity measures the level of system activity determined by jump statistics, $\mathcal{B}(\tau)$ evaluates the activity of open quantum dynamics within $[0,\tau]$. In quantum dynamics, state changes can occur even without a jump, owing to the coherent dynamics. Thus, $\mathcal{B}(t)$ includes contributions from jump and coherent dynamics.
As mentioned, the classical case can be covered by considering specific $H_S$ and $L_m$ in the Lindblad equation [Eq.~\eqref{eq:superop_def}] in which
$\mathcal{B}(t)$ reduces to $\mathcal{A}(t)$ for the classical case. 
On the other hand, for $L_m=0$ for all $m$, the Lindblad equation in Eq.~\eqref{eq:superop_def} becomes the closed quantum dynamics driven by the Hamiltonian $H_S$. 
Details on the quantum dynamical activity are presented in Ref.~\cite{Supp:2024:ConcenIneq}.

\textit{Results.---}Because the dynamical activities in classical and quantum models measure the degree of activity, these measures could have a quantitative relationship with the probability distribution $P(N(\tau))$.
We claim that the dynamical activity and probability $P(N(\tau)=0)$ are related by inequalities. 
Specifically, for $0\le(1/2)\int_{0}^{\tau}\sqrt{\mathcal{B}(t)}/t\,dt\le\pi/2$, the thermodynamic concentration inequality holds:
  \begin{align}
    \cos\left[\frac{1}{2}\int_{0}^{\tau}\frac{\sqrt{\mathcal{B}(t)}}{t}dt\right]^{2}\le P(N(\tau)=0).
    \label{eq:main_result_quantum}
\end{align}
Equation~\eqref{eq:main_result_quantum} is the key component for obtaining tradeoff relations through other concentration inequalities
and constitutes the first result of this Letter. 
A proof of Eq.~\eqref{eq:main_result_quantum} is presented in Ref.~\cite{Supp:2024:ConcenIneq}. 
As the quantum dynamical activity, denoted by $\mathcal{B}(t)$, increases, the probability $P(N(\tau)=0)$ decreases. Consequently, the left-hand side of Eq.~\eqref{eq:main_result_quantum} decreases, which aligns with our intuitive understanding.
Equation~\eqref{eq:main_result_quantum} can be equivalently expressed using $\sin\left[(1/2)\int_{0}^{\tau}\sqrt{\mathcal{B}(t)}/t\,dt\right]^{2}\ge P(|N(\tau)|>0)$, which estimates the probability that $|N(\tau)|$ deviates from $0$. 
In the classical limit, $\mathcal{B}(\tau)$ is simplified to $\mathcal{A}(\tau)$ to directly derive the classical case of Eq.~\eqref{eq:main_result_quantum}.
For a time-independent classical Markov process and $0\le(1/2)\int_{0}^{\tau}\sqrt{\mathcal{A}(t)}/t\,dt\le\pi/2$, we obtain
\begin{align}
    \cos\left[\frac{1}{2}\int_{0}^{\tau}\frac{\sqrt{\mathcal{A}(t)}}{t}dt\right]^{2}\le P(N(\tau)=0).
    \label{eq:main_result_classical}
\end{align}
Any relations that hold for $\mathcal{B}(\tau)$ should hold for $\mathcal{A}(\tau)$ when considering the classical limit. 
For classical Markov processes, a different concentration inequality holds. For a time-independent classical Markov process, we obtain
  \begin{align}
    e^{-\mathfrak{a}(0)\tau}\le P(N(\tau)=0),
    \label{eq:main_result_classical_stronger}
\end{align}
where $\mathfrak{a}(t)$ denotes the rate of dynamical activity [Eq.~\eqref{eq:mathfrak_a_def}].
A proof of Eq.~\eqref{eq:main_result_classical_stronger} is presented in Ref.~\cite{Supp:2024:ConcenIneq}. 
A notable advantage of Eq.~\eqref{eq:main_result_classical_stronger} over Eq.~\eqref{eq:main_result_classical} is that it holds for any $\tau > 0$. 
In general, it is unknown which of Eqs.~\eqref{eq:main_result_classical} and \eqref{eq:main_result_classical_stronger} is tighter. However, in the steady-state condition, Eq.~\eqref{eq:main_result_classical_stronger} is tighter (see Ref.~\cite{Supp:2024:ConcenIneq}).
Entropy production plays a fundamental role in thermodynamic tradeoff relations. Entropy production measures the irreversibility of a process, while dynamical activity indicates its inherent timescale.
In classical Markov processes, there is an inequality relationship between entropy production and classical dynamical activity \cite{Nishiyama:2023:EPUpperBound}. This inequality indicates that positive entropy production necessitates positive classical dynamical activity; however, the reverse is not always the case.

In the previous section, we derived concentration inequalities in the context of classical and quantum thermodynamics.
The obtained concentration inequalities provide bounds on the number of jumps and assist in deriving various tradeoff relations.
We obtain tradeoff relations that can be derived from Eqs.~\eqref{eq:main_result_quantum}--\eqref{eq:main_result_classical_stronger} by combining them with the other concentration inequalities. 
The most well-known concentration inequality is the Markov inequality [Eq.~\eqref{eq:Markov_inequality}], considered a fundamental tool because of its simplicity and generality. It does not require knowledge of the probability distribution, only its mean.
Using the Markov inequality, we can derive an upper bound for $\mathbb{E}[|N(\tau)|]$.
Let $N_\mathrm{max}$ be the maximum of $|N(\tau)|$. 
Then, for $0\le(1/2)\int_{0}^{\tau}\sqrt{\mathcal{B}(t)}/t\,dt\le\pi/2$, the following relation can be derived from Eq.~\eqref{eq:main_result_quantum}:
\begin{align}
    \mathbb{E}[|N(\tau)|]&\leq N_{\max}\sin\left[\frac{1}{2}\int_{0}^{\tau}\frac{\sqrt{\mathcal{B}(t)}}{t}dt\right]^{2}.\label{eq:thermo_Markov_ineq}
\end{align}
For a classical Markov process, the following relation can be derived from Eq.~\eqref{eq:main_result_classical_stronger}:
\begin{align}
\mathbb{E}[|N(\tau)|]&\leq N_{\max}\left(1-e^{-\mathfrak{a}(0)\tau}\right).
    \label{eq:thermo_Markov_ineq_mathfraka}
\end{align}
Proofs of Eqs.~\eqref{eq:thermo_Markov_ineq} and \eqref{eq:thermo_Markov_ineq_mathfraka} are based on the application of Eqs.~\eqref{eq:main_result_quantum} and \eqref{eq:main_result_classical_stronger} to the Markov inequality, as shown in Ref.~\cite{Supp:2024:ConcenIneq}. 
Equations~\eqref{eq:thermo_Markov_ineq} and \eqref{eq:thermo_Markov_ineq_mathfraka} constitute the second result of this Letter. 
Given the general observables satisfying Eq.~\eqref{eq:Ovarnothing_def},
Eqs.~\eqref{eq:thermo_Markov_ineq} and \eqref{eq:thermo_Markov_ineq_mathfraka} hold for any quantum and classical Markov processes, respectively.
In stochastic thermodynamics, when observable such as the number of transported particles or jumps are concerned, $N_{\mathrm{max}}$ is infinite and thus Eqs.~\eqref{eq:thermo_Markov_ineq} and \eqref{eq:thermo_Markov_ineq_mathfraka} become uninformative.
However, in the analysis of the correlation function, $N_{\mathrm{max}}$ remains finite in which we can derive upper bounds for the correlation function, as demonstrated in Ref.~\cite{Hasegawa:2024:TCI}.
Moreover, from the correlation bounds, we can obtain an upper bound for the response against perturbation in a classical Markov process using the linear response theory. 
This is discussed in detail in Ref.~\cite{Supp:2024:ConcenIneq}.

Probability theory and statistics use the integral probability metric to measure the distance between two probability distributions. It includes several well-known distance metrics,
such as the Wasserstein-1 distance and the total variation distance.
Speed limits in classical Markov processes can be derived using the integral probability metric using Eq.~\eqref{eq:thermo_Markov_ineq_mathfraka}. 
Classical speed limits using the integral probability metric were recently considered in Ref.~\cite{Kwon:2023:XTUR},
which proposed deriving several uncertainty relations in a unified framework. 
The integral probability metric between two probability distributions $\mathfrak{P}(X)$ and $\mathfrak{Q}(Y)$ is defined as follows:
\begin{align}
    D_{\mathcal{F}}(\mathfrak{P},\mathfrak{Q})\equiv\max_{f\in\mathcal{F}}\left|\mathbb{E}_{\mathfrak{P}}\left[f(X)\right]-\mathbb{E}_{\mathfrak{Q}}\left[f(Y)\right]\right|,
    \label{eq:DF_def}
\end{align}
where $\mathbb{E}_{\mathfrak{P}}[\bullet]$ denotes the expectation with respect to $\mathfrak{P}(X)$. 
Here, $\mathcal{F}$ denotes a class of real-valued functions and their choice specifies the type of distance.
For example, Eq.~\eqref{eq:DF_def} expresses the total variation distance, where $\mathcal{F}$ is a set of indicator functions that return either $0$ or $1$.  
From Eq.~\eqref{eq:traj_classical_def},
we consider the following observable $N(\tau)$:
\begin{align}
    N(\tau) = f(X(\tau)) - f(X(0)),
    \label{eq:N_for_IPM}
\end{align}
which satisfies Eq.~\eqref{eq:Ovarnothing_def}.
Subsequently, the expectation is
\begin{align}
    \ensuremath{\mathbb{E}[N(\tau)]}=\mathbb{E}_{\mathbf{P}(\tau)}\left[f(X(\tau))\right]-\mathbb{E}_{\mathbf{P}(0)}\left[f(X(0))\right].
    \label{eq:N_to_IPM}
\end{align}
From Eq.~\eqref{eq:N_to_IPM}, $D_{\mathcal{F}}(\mathbf{P}(0),\mathbf{P}(\tau))=\max_{f\in\mathcal{F}}\left|\mathbb{E}[N(\tau)]\right|$. 
Substituting Eq.~\eqref{eq:N_to_IPM} into Eq.~\eqref{eq:thermo_Markov_ineq_mathfraka}, the following result is obtained:
\begin{align}
    D_{\mathcal{F}}(\mathbf{P}(\tau),\mathbf{P}(0))\leq F_{\max}\left(1-e^{-\mathfrak{a}(0)\tau}\right),\label{eq:IMP_SL_stronger}
\end{align}
where 
\begin{align}
    F_{\max}\equiv\max_{X_{1},X_{2}\in\mathfrak{B}}|f_{\max}(X_{1})-f_{\max}(X_{2})|,
    \label{eq:fmax_def}
\end{align}
where $f_{\max}\equiv\mathrm{argmax}_{f\in\mathcal{F}}\left|\mathbb{E}[N(\tau)]\right|$. 
As mentioned, depending on the choice of $\mathcal{F}$, Eq.~\eqref{eq:IMP_SL_stronger} becomes a bound for a different distance function.
We discuss only the speed limits of classical Markov processes. 
However, their quantum generalization is not trivial
because the quantum distance measure does not permit an integral probability metric representation.  

Next, we relate the probability $P(N(\tau)=0)$ to the norm of $N(\tau)$ using another concentration inequality, 
namely Petrov inequality (Eq.~(9) in \cite{Valentin:2007:TailProb}):
\begin{align}
    P(|X|>b)\ge\frac{\left(\mathbb{E}[|X|^{r}]-b^{r}\right)^{s/(s-r)}}{\mathbb{E}[|X|^{s}]^{r/(s-r)}},
    \label{eq:Paley_Zygmund_ineq}
\end{align}
where $0<r<s$, $b \ge 0$, and the condition $b^{r}\le\mathbb{E}[|X|^{r}]$ should be satisfied. 
The well-known Paley-Zygmund inequality \cite{Paley:1932:PaleyZygmundIneq} can be obtained by substituting $r=1$, $s=2$, and $b = \lambda \mathbb{E}[|X|]$ in Eq.~\eqref{eq:Paley_Zygmund_ineq}, where $0\le \lambda \le 1$. 
Using Eqs.~\eqref{eq:main_result_quantum}, \eqref{eq:main_result_classical_stronger}, and \eqref{eq:Paley_Zygmund_ineq}, the thermodynamic uncertainty relation can be generalised. 
For $0\le(1/2)\int_{0}^{\tau}\sqrt{\mathcal{B}(t)}/t\,dt\le\pi/2$ and $0<r<s$, the following relation holds:
\begin{align}
\frac{\mathbb{E}[|N(\tau)|^{s}]^{r/(s-r)}}{\mathbb{E}[|N(\tau)|^{r}]^{s/(s-r)}}&\ge\sin\left[\frac{1}{2}\int_{0}^{\tau}\frac{\sqrt{\mathcal{B}(t)}}{t}dt\right]^{-2}.\label{eq:TUR_main}
\end{align}
Moreover, for the classical Markov process, we obtain
\begin{align}
\frac{\mathbb{E}[|N(\tau)|^{s}]^{r/(s-r)}}{\mathbb{E}[|N(\tau)|^{r}]^{s/(s-r)}}&\ge\frac{1}{1-e^{-\mathfrak{a}(0)\tau}}.
\label{eq:TUR_classical_stronger}
\end{align}
A proof of Eqs.~\eqref{eq:TUR_main} and \eqref{eq:TUR_classical_stronger} is a direct application of Eqs.~\eqref{eq:main_result_quantum} and \eqref{eq:main_result_classical_stronger} to Eq.~\eqref{eq:Paley_Zygmund_ineq}. 
Equations~\eqref{eq:TUR_main} and \eqref{eq:TUR_classical_stronger} hold for arbitrary time-independent quantum and classical Markov processes with arbitrary initial states, respectively. 
Equations~\eqref{eq:TUR_main} and \eqref{eq:TUR_classical_stronger} are the third result of this Letter.
For the random variable $X$, 
we define the $p$-norm as $\left\Vert X\right\Vert _{p}\equiv\mathbb{E}[|X|^{p}]^{1/p}$ for $p>1$, which is widely used in statistics. Substituting $r=1$ and $s = p>1$ in Eq.~\eqref{eq:TUR_main}, we obtain the bound for the $p$-norm as follows:
\begin{align}
    \frac{\left\Vert N(\tau)\right\Vert _{p}}{\left\Vert N(\tau)\right\Vert _{1}}\ge\sin\left[\frac{1}{2}\int_{0}^{\tau}\frac{\sqrt{\mathcal{B}(t)}}{t}dt\right]^{-\frac{2(p-1)}{p}},
    \label{eq:p_norm_TUR}
\end{align}
which holds for $0\le(1/2)\int_{0}^{\tau}\sqrt{\mathcal{B}(t)}/t\,dt\le\pi/2$.
Similarly, Eq.~\eqref{eq:TUR_classical_stronger} yields
\begin{align}
\frac{\left\Vert N(\tau)\right\Vert _{p}}{\left\Vert N(\tau)\right\Vert _{1}}\ge\frac{1}{\left(1-e^{-\mathfrak{a}(0)\tau}\right)^{\frac{p-1}{p}}}.
    \label{eq:p_norm_TUR_stronger_classical}
\end{align}
For $p=2$, Eqs.~\eqref{eq:p_norm_TUR} and \eqref{eq:p_norm_TUR_stronger_classical} recover the thermodynamic uncertainty relations for the mean and variance of $N(\tau)$:
\begin{align}
    \frac{\mathrm{Var}[|N(\tau)|]}{\mathbb{E}[|N(\tau)|]^{2}}&\ge\tan\left[\frac{1}{2}\int_{0}^{\tau}\frac{\sqrt{\mathcal{B}(t)}}{t}dt\right]^{-2},\label{eq:TUR1}
\end{align}
and
\begin{align}
    \frac{\mathrm{Var}[|N(\tau)|]}{\mathbb{E}[|N(\tau)|]^{2}}&\ge\frac{1}{e^{\mathfrak{a}(0)\tau}-1},
    \label{eq:TUR1_classical_stronger}
\end{align}
where $\mathrm{Var}[X] \equiv \mathbb{E}[X^2]-\mathbb{E}[X]^2$ denotes the variance of $X$. 
The absolute value symbol on the left side of Eqs.~\eqref{eq:TUR1} and \eqref{eq:TUR1_classical_stronger} can be removed
given $\mathbb{E}[N(\tau)] > 0$
because
$\mathrm{Var}[N(\tau)]/\mathbb{E}[N(\tau)]^{2}\ge\mathrm{Var}[|N(\tau)|]/\mathbb{E}[|N(\tau)|]^{2}$, which follows from $\mathbb{E}[|N(\tau)|]\ge\mathbb{E}[N(\tau)]$ and $\mathrm{Var}[|N(\tau)|]\le\mathrm{Var}[N(\tau)]$. 
Note that Eq.~\eqref{eq:TUR1} was reported in Ref.~\cite{Hasegawa:2023:BulkBoundaryBoundNC}. 
In addition, in the limit of $p\to \infty$, Eqs.~\eqref{eq:p_norm_TUR} and \eqref{eq:p_norm_TUR_stronger_classical} reduce to Eqs.~\eqref{eq:thermo_Markov_ineq} and \eqref{eq:thermo_Markov_ineq_mathfraka}, respectively, since $\left\Vert X\right\Vert _{\infty}=\max|X|$. 

Our focus has been on the theoretical aspects of the bounds. To validate the obtained bounds, we perform numerical simulations on the trade-off relations. For more detailed information on these numerical simulations, refer to Ref.~\cite{Supp:2024:ConcenIneq}.

\textit{Discussion.---}This Letter derived tradeoff relations using concentration inequalities, typically discussed in terms of information inequalities, such as the Cram\'er-Rao inequality. 
Although the tradeoff relations obtained by the information and concentration inequalities differ, they can be understood in the context of a thermodynamic uncertainty relation.
The following thermodynamic uncertainty relation is known
\cite{Hasegawa:2023:BulkBoundaryBoundNC}:
\begin{align}
\left(\frac{\sqrt{\mathrm{Var}[N(t_{2})]}+\sqrt{\mathrm{Var}[N(t_{1})]}}{\mathbb{E}[N(t_{2})]-\mathbb{E}[N(t_{1})]}\right)^{2}\geq\tan\left[\frac{1}{2}\int_{t_{1}}^{t_{2}}\frac{\sqrt{\mathcal{B}(t)}}{t}dt\right]^{-2},
    \label{eq:interpolating_TUR}
\end{align}
where $0 < t_1 < t_2$. 
First, let us consider short-time limits, that is, $t_1$ and $t_2$ are infinitesimally close. 
When $t_2 = \tau$ and $t_1 = \tau - \epsilon$, where $\epsilon$ is infinitesimally small, Eq.~\eqref{eq:interpolating_TUR} reduces to
\begin{align}
    \frac{\mathrm{Var}[N(\tau)]}{\tau^{2}\left(\partial_{\tau}\mathbb{E}[N(\tau)]\right)^{2}}\geq\frac{1}{\mathcal{B}(\tau)}.
    \label{eq:TUR_diff}
\end{align}
which can be derived via the Cram\'er--Rao inequality \cite{Hasegawa:2020:QTURPRL}. 
The classical case of Eq.~\eqref{eq:TUR_diff} was derived in Refs.~\cite{Garrahan:2017:TUR,Terlizzi:2019:KUR}, where the quantum dynamical activity $\mathcal{B}(\tau)$ is replaced by the classical dynamical activity $\mathcal{A}_\mathrm{cl}(\tau)$. 
In contrast, when $t_1$ and $t_2$ are far apart, that is, $t_2 = \tau$ and $t_1 = 0$, Eq.~\eqref{eq:interpolating_TUR} agrees with Eq.~\eqref{eq:TUR1}, which was derived from the concentration inequality in this Letter. 
From this observation, the short-term case corresponds to information inequality, whereas the long-term case corresponds to concentration inequality.
This shows that the Cram\'er-Rao inequality uses information only around $\tau$ whereas the concentration inequality uses information within the range of $[0,\tau]$.
This implies that concentration and information inequalities exploit the complementary aspects of the dynamics. 

The approach presented in this Letter lays the foundation for uncovering new uncertainty relations by combining the thermodynamic concentration inequalities with other prevalent ones.
In this Letter, we focused on the probability $P(N(\tau)=0)$. 
We can consider, for example, the possibility of examining a probability distribution other than $P(N(\tau)=0)$. 
Similarly, we could also take into account thermodynamic costs not restricted to dynamical activity. 
We reserve these generalizations for future research.

\begin{acknowledgments}

This work was supported by JSPS KAKENHI Grant Number JP23K24915.

\end{acknowledgments}

\end{document}


\title{Supplementary Material for\\ ``Thermodynamic Concentration Inequalities and Tradeoff Relations''}
\author{Yoshihiko Hasegawa}
\email{hasegawa@biom.t.u-tokyo.ac.jp}
\affiliation{Department of Information and Communication Engineering, Graduate
School of Information Science and Technology, The University of Tokyo,
Tokyo 113-8656, Japan}

\author{Tomohiro Nishiyama}
\email{htam0ybboh@gmail.com}
\affiliation{Independent Researcher, Tokyo 206-0003, Japan}

\maketitle
This supplementary material describes the calculations introduced in the main text. The numbers of the equations and the figures are prefixed with S (e.g., Eq.~(S1) or Fig.~S1). Numbers without this prefix (e.g., Eq.~(1) or Fig.~1) refer to items in the main text.

\section{Lindblad equation}
\subsection{Continuous measurement}

We review the basics of continuous measurement formalism in open quantum dynamics (see Ref.~\cite{Landi:2023:CurFlucReviewPRXQ} for a comprehensive review). 
Let us define the non-Hermitian effective Hamiltonian:
\begin{align}
    H_{\mathrm{eff}}\equiv H_{S}-\frac{i}{2}\sum_{m=1}^{N_{C}}L_{m}^{\dagger}L_{m}.
    \label{eq:Heff_def}
\end{align}
By using Eq.~\eqref{eq:Heff_def}, we can express Eq.~\superopUdef{} in the following manner at the $O(dt)$ order:
\begin{align}
\rho_{S}(t+dt)&=\rho_{S}(t)-idt\left(H_{\mathrm{eff}}\rho_{S}(t)-\rho_{S}(t)H_{\mathrm{eff}}^{\dagger}\right)+dt\sum_{m=1}^{N_{C}}L_{m}\rho_{S}(t)L_{m}^{\dagger}.
    \label{eq:rhoS_dt}
\end{align}
From Eq.~\eqref{eq:rhoS_dt}, it can be seen that the dynamics is described by the continuous state transitions by the non-Hermitian operator $H_{\mathrm{eff}}$, and the discontinuous state changes by the jump operator $L_m$. 
Because a completely positive and trace-preserving map can be represented using the Kraus representation, the time evolution in Eq.~\eqref{eq:rhoS_dt} can be described using the Kraus representation:
\begin{align}
    \rho_{S}(t+dt)=\sum_{m=0}^{N_{C}}V_{m}\rho_{S}(t)V_{m}^{\dagger},
    \label{eq:Kraus_repr}
\end{align}
where $V_m$ is the Kraus operator defined as
\begin{align}
    V_{0}&=\mathbb{I}_{S}-idtH_{\mathrm{eff}},\label{eq:Kraus_V0_def}\\
    V_{m}&=\sqrt{dt}L_{m}\hspace*{1em}(1\le m\le N_{C}).\label{eq:Kraus_Vm_def}
\end{align}
By virtue of the completeness relation, $V_m$ should satisfy $\sum_{m=0}^{N_{C}}V_{m}^{\dagger}V_{m}=\mathbb{I}_{S}$, where $\mathbb{I}_S$ is the identity operator of the system. 

The Kraus representation in Eq.~\eqref{eq:Kraus_repr} allows us to identify the dynamics of the Lindblad equation as a consequence of the measurement, 
where $m$ denotes the measurement output
($m=0$ corresponds to the no-jump output).
Specifically, when the output is $m$, the post-measurement state becomes $\rho_{S}(t+dt)=V_{m}\rho_{S}(t)V_{m}^{\dagger}/\mathrm{Tr}_S[V_{m}\rho_{S}(t)V_{m}^{\dagger}]$. 
Suppose that the dynamics starts at $t=0$ and ends at $t=\tau$. 
Let $J$ be a sufficiently large integer, and $dt = \tau / J$. 
Applying Eq.~\eqref{eq:Kraus_repr} repeatedly, $\rho_S(\tau)$ is represented by
\begin{align}
\rho_{S}(\tau)=\sum_{\mathbf{m}}\mathcal{V}_{\mathbf{m}}\rho_{S}(0)\mathcal{V}_{\mathbf{m}}^{\dagger},
    \label{eq:Kraus_tau}
\end{align}
where $\mathbf{m} \equiv [m_{J-1},\cdots,m_{0}] \in \{0,1,\cdots,N_C\}^{J}$
denotes the measurement record and $\mathcal{V}_{\mathbf{m}}\equiv V_{m_{J-1}}\cdots V_{m_{0}}$. 
The variable $\mathbf{m}$ represents whether a jump occurs in each short interval $dt$. 
This information can also be expressed by specifying the time $t_k$ at which each jump occurs and the type of jump $m_k$ ($1 \le m_k \le N_C$.  $m_k=0$, which corresponds to no jump, is excluded), which is essentially $\zeta_\tau$ as outlined in Eq.~\trajUzetaUdef{}.

\subsection{Matrix product state representation}

Next, we introduce the matrix product state formalism in the continuous measurement. 
From the Kraus representation in Eq.~\eqref{eq:Kraus_tau}, the following continuous matrix product state representation is considered:
\begin{align}
    \ket{\Phi(\tau)}&=\sum_{\mathbf{m}}\mathcal{V}_{\mathbf{m}}\ket{\psi_{S}(0)}\otimes\ket{\mathbf{m}},
    \label{eq:Psi_tau_def}
\end{align}
where $\ket{\psi_S(0)}$ is the initial state vector ($\rho_S(0)=\ket{\psi_S(0)}\bra{\psi_S(0)}$). 
Considering $dt \to 0$ limit in Eq.~\eqref{eq:Psi_tau_def}, while keeping $\tau$ constant, 
$\ket{\Psi(t)}$ can be represented as a continuous matrix product state \cite{Verstraete:2010:cMPS,Osborne:2010:Holography}:
\begin{align}
    \ket{\Phi(t)}=\mathcal{U}(t)\ket{\psi_{S}(0)}\otimes\ket{\mathrm{vac}},
    \label{eq:cMPS_def}
\end{align}
where $\mathcal{U}(t)$ is the operator defined by
\begin{align}
    \mathcal{U}(t)=\mathbb{T}e^{-i\int_{0}^{t}dt^{\prime}\,(H_{\mathrm{eff}}\otimes\mathbb{I}_{F}+\sum_{m}iL_{m}\otimes\phi_{m}^{\dagger}(t^{\prime}))}.
    \label{eq:mathcalV_def}
\end{align}
Here, $\mathbb{T}$ denotes the time-ordering operator, $\mathbb{I}_F$ is the identity operator in the field, $\phi_m(t)$ is a field operator that satisfies the canonical commutation relation $[\phi_{m}(t),\phi_{m^{\prime}}^{\dagger}(t^{\prime})]=\delta_{mm^{\prime}}\delta(t-t^{\prime})$, and 
$|\mathrm{vac}\rangle$ denotes the vacuum state. 
Within this setting, the continuous matrix product state captures all continuous measurement information by generating particles by applying $\phi_m^{\dagger}(t)$ to the vacuum state. The benefit of using the continuous matrix product state for continuous measurements is its ability to embed jump events and the system state into a pure state $\ket{\Phi(t)}$.

\subsection{Trajectory}

The probability $\mathcal{P}(\zeta_\tau)$ of measuring $\zeta_\tau$ as defined in Eq.~\trajUzetaUdef{} is equivalent to the probability of measuring $\mathbf{m} = [m_{J-1}, \cdots, m_0]\in \{0,1,\cdots,N_C\}^J$ and $s$.
Using the state $\ket{\Phi(\tau)}$, 
this relationship allows us to express the probability as follows:
\begin{align}
\mathcal{P}(\zeta_{\tau}) = P(\mathbf{m}, s) = \braket{\Phi(\tau)|\Xi_{s} \otimes \Pi_{\mathbf{m}}|\Phi(\tau)},
\label{eq:P_zetatau_def}
\end{align}
where $\Pi_{\mathbf{m}} \equiv \ket{\mathbf{m}}\bra{\mathbf{m}}$ is the projector associated with the measurement of $\mathbf{m}$ in the environment, and $\Xi_s$ is the projector corresponding to the system output $s$.

In the quantum case, even if we know the final state of the system, it does not completely determine the system's time evolution. 
In contrast, in the classical case, knowing the trajectory $\zeta_\tau$ fully specifies the system's time evolution.
Therefore, the trajectory $\zeta_\tau$ can be represented by
\begin{align}
    \zeta_\tau = \{X(t) | 0 \le t \le \tau\},
    \label{eq:zetatau_classical}
\end{align}
where
$X(t) \in \mathfrak{B}$ is the state at time $t$ as defined in the main text.
Equation~\eqref{eq:zetatau_classical} corresponds to Eq.~\trajUclassicalUdef{} in the main text. 

\section{Bound on the fidelity}

We review the upper bound of fidelity in Ref.~\cite{Hasegawa:2023:BulkBoundaryBoundNC}, which is used for the thermodynamic concentration inequalities. 

Let $\ket{\psi(t)}$ be a general state vector at time $t$ of general quantum dynamics. 
Let $\mathcal{J}(t)$ be the quantum Fisher information with respect to $t$, given by
\begin{align}
    \mathcal{J}(t)\equiv4\left[\braket{\partial_{t}\psi(t)|\partial_{t}\psi(t)}-\left|\braket{\partial_{t}\psi(t)|\psi(t)}\right|^{2}\right].
    \label{eq:QFI_def}
\end{align}
Then, the following inequality holds \cite{Uhlmann:1992:BuresGeodesic,Taddei:2013:QSL}:
\begin{align}
    \frac{1}{2}\int_{t_{1}}^{t_{2}}dt\sqrt{\mathcal{J}(t)}\geq\arccos\left|\braket{\psi(t_{2})|\psi(t_{1})}\right|,
    \label{eq:QFI_fidelity_ineq}
\end{align}
where the left-hand side represents the length of the path traversed through time evolution and the right-hand side represents the distance (i.e., the geodesic distance) between $\ket{\psi(t_1)}$ and $\ket{\psi(t_2)}$.

The continuous matrix product state given by Eqs.~\eqref{eq:cMPS_def} and \eqref{eq:mathcalV_def} is inconvenient to evaluate the fidelity between two states at different times $t_1$ and $t_2$, $\braket{\Phi(t_2)|\Phi(t_1)}$. 
This is because evaluating the fidelity between two continuous matrix product states, denoted by $\left\langle\Phi\left(t_2\right) | \Phi\left(t_1\right)\right\rangle$, where $t_1 \neq t_2$ is unfeasible because of the different integration ranges for $\left|\Phi\left(t_1\right)\right\rangle$ and $\left|\Phi\left(t_2\right)\right\rangle$ as shown in Eqs.~\eqref{eq:cMPS_def} and \eqref{eq:mathcalV_def}. Consequently, in our study, we adopt the scaled continuous matrix product state \cite{Hasegawa:2023:BulkBoundaryBoundNC}:
\begin{align}
    \ket{\Psi(t)}\equiv\mathcal{V}(\theta=t/\tau)\ket{\psi_{S}(0)}\otimes\ket{\mathrm{vac}},
    \label{eq:scaled_cMPS}
\end{align}
where $\theta \equiv t/ \tau$ is the scaled time and
\begin{align}
    \mathcal{V}(\theta)\equiv\mathbb{T}e^{\int_{0}^{\tau}ds\,(-i\theta H_{\mathrm{eff}}\otimes\mathbb{I}_{\mathrm{F}}+\sqrt{\theta}\sum_{m}L_{m}\otimes\phi_{m}^{\dagger}(s))}.
    \label{eq:mathcalV_theta_def}
\end{align}
Using this representation, 
we can consider the 
time evolution of $\ket{\Psi(t)}$ to evaluate the fidelity at different times because the integration range in Eq.~\eqref{eq:mathcalV_theta_def} does not depend on $t$.

Consider the geometric relation given by Eq.~\eqref{eq:QFI_fidelity_ineq} for the scaled continuous matrix product state $\ket{\Psi(t)}$. 
Then, we have the following relation \cite{Hasegawa:2023:BulkBoundaryBoundNC}:
\begin{align}
    \frac{1}{2}\int_{0}^{\tau}\frac{\sqrt{\mathcal{B}(t)}}{t}dt\geq\arccos\left|\braket{\Psi(\tau)|\Psi(0)}\right|.
    \label{eq:QDA_Bures_angle}
\end{align}
where $\mathcal{B}(t)$ is defined as
\begin{align}
    \mathcal{B}(t) =4t^{2}\left(\braket{\partial_{t}\Psi(t)|\partial_{t}\Psi(t)}-\left|\braket{\Psi(t)|\partial_{t}\Psi(t)}\right|^{2}\right),
    \label{eq:QDA_def}
\end{align}
which is the quantum dynamical activity. 
Although the right-hand side includes $\tau$, it can be demonstrated that $\mathcal{B}(t)$ does not depend on the choice of $\tau$.

\section{Quantum dynamical activity}

Quantum dynamical activity is important in tradeoff relations \cite{Hasegawa:2020:QTURPRL,Hasegawa:2023:BulkBoundaryBoundNC,
Nakajima:2023:SLD,Nishiyama:2024:ExactQDAPRE,Nishiyama:2024:OpenQuantumRURJPA}. 
For example, for the counting observable $N(\tau)$ (cf. Eq.~\MUdef{}) that counts the number of jump events within $[0,\tau]$ under steady-state conditions, the following quantum thermodynamic uncertainty relation holds \cite{Hasegawa:2020:QTURPRL}:
\begin{align}
    \frac{\mathrm{Var}[N(\tau)]}{\mathbb{E}[N(\tau)]^{2}}\ge\frac{1}{\mathcal{B}(\tau)},
    \label{eq:quantum_TUR_original}
\end{align}
which was derived via the quantum Cram\'er-Rao inequality. 
The quantum dynamical activity also plays an important role in a quantum speed limit. 
Let $\mathcal{L}_{D}(\rho(0),\rho(\tau))\equiv\arccos\left[\sqrt{\mathrm{Fid}(\rho(0),\rho(\tau))}\right]$ be the Bures angle between the initial state $\rho_S(0)$ and the final state $\rho_S(\tau)$, where $\mathrm{Fid}(\rho_{1},\rho_{2})\equiv\left(\mathrm{Tr}\sqrt{\sqrt{\rho_{1}}\rho_{2}\sqrt{\rho_{1}}}\right)^{2}$ is the fidelity. 
Then, the following quantum speed limit is known to hold \cite{Hasegawa:2023:BulkBoundaryBoundNC}:
\begin{align}
    \mathcal{L}_D(\rho_S(0), \rho_S(\tau)) \le \frac{1}{2} \int_0^\tau d t \frac{\sqrt{\mathcal{B}(t)}}{t}.
    \label{eq:QST_by_Bt}
\end{align}

The exact expression for $\mathcal{B}(t)$ in the quantum dynamical activity defined in Eq.~\eqref{eq:QDA_def} was recently obtained \cite{Nishiyama:2024:ExactQDAPRE} (in Ref.~\cite{Nakajima:2023:SLD}, an intermediate expression was derived):
\begin{align}
    \mathcal{B}(\tau)=\mathcal{A}(\tau)+\mathcal{C}(\tau),
    \label{eq:QDA_exact_solution}
\end{align}
where $\mathcal{A}(\tau)$ denotes the classical dynamical activity [Eq.~\classicalUDAUdefII{}] and 
$\mathcal{C}(t)$ is the coherent dynamics contribution given by
\begin{align}
\mathcal{C}(\tau)&\equiv8\int_{0}^{\tau}ds_{1}\int_{0}^{s_{1}}ds_{2}\mathrm{Re}[\mathrm{Tr}_{S}\{H_{\mathrm{eff}}^{\dagger}\check{H}_{S}(s_{1}-s_{2})\rho_{S}(s_{2})\}]-4\left(\int_{0}^{\tau}ds\mathrm{Tr}_{S}\left[H_{S}\rho_{S}(s)\right]\right)^{2}.
    \label{eq:QDA_Bq_def}
\end{align}
Here, $\check{H}_S(t) \equiv e^{\mathcal{L}^{\dagger} t} H_S$ is interpreted as the Hamiltonian $H_S$ in the Heisenberg picture, where $\mathcal{L}^\dagger$ is the adjoint superoperator
\begin{align}
    \mathcal{L}^{\dagger}\mathcal{O}\equiv i\left[H_{S},\mathcal{O}\right]+\sum_{m=1}^{N_{C}}\mathcal{D}^{\dagger}[L_{m}]\mathcal{O},
    \label{eq:mathcalL_dagger_def}
\end{align}
 and $\mathcal{D}^\dagger$ is the adjoint dissipator:
\begin{align}
    \mathcal{D}^{\dagger}[L]\mathcal{O}\equiv L^{\dagger}\mathcal{O}L-\frac{1}{2}\{L^{\dagger}L,\mathcal{O}\},
    \label{eq:mathcalD_dagger_def}
\end{align}
for an operator $\mathcal{O}$. 
As indicated previously, the classical dynamical activity quantifies the degree of activity of the system and can be evaluated using jump statistics. 
The quantum dynamical activity also quantifies the system's activity for quantum Markov processes. 
However, state changes can occur even when there is no jump in the quantum Markov processes owing to the coherent dynamics induced by the Hamiltonian $H_S$; $\mathcal{C}(\tau)$ in Eq.~\eqref{eq:QDA_exact_solution} reflects such contributions. 
Apparently, for the classical limit, where $H_S=0$, $\mathcal{B}(\tau)$ reduces to $\mathcal{A}(\tau)$. 
When we replace $\mathcal{B}(\tau)$ in Eq.~\eqref{eq:quantum_TUR_original} with $\mathcal{A}(\tau)$, it becomes identical to the bound reported in Ref.~\cite{Garrahan:2017:TUR}.
Therefore, the quantum thermodynamic uncertainty relation of Eq.~\eqref{eq:quantum_TUR_original} includes the classical counterpart \cite{Garrahan:2017:TUR} as the classical limit. 
Now, let us examine the opposite limit: closed quantum dynamics. This case can be modeled by setting all Lindblad operators $L_m$ to zero. Under this condition, the Lindblad equation [Eq.~\superopUdef{} in the main text] reduces to a description of closed quantum dynamics governed by the system Hamiltonian $H_S$. Then the quantum dynamical activity becomes \cite{Nishiyama:2024:ExactQDAPRE}
\begin{align}
    \mathcal{B}(\tau)=4\tau^{2}\left(\mathrm{Tr}_{S}\left[H_{S}^{2}\rho_{S}\right]-\mathrm{Tr}_{S}\left[H_{S}\rho_{S}\right]^{2}\right) = 4\tau^2 \dblbrace{H_S}^2,
    \label{eq:Bt_closed}
\end{align}
which is the variance of $H_S$ multiplied by $4\tau^2$. 
When substituting Eq.~\eqref{eq:Bt_closed} into Eq.~\eqref{eq:QST_by_Bt}, we obtain
\begin{align}
    \mathcal{L}_{D}(\rho_{S}(0),\rho_{S}(\tau))\le\tau\dblbrace{H_{S}}.
    \label{eq:MT_bound}
\end{align}
Since $\arccos 0 = \pi/2$, Eq.~\eqref{eq:MT_bound} is equivalent to the Mandelstam-Tamm bound \cite{Mandelstam:1945:QSL}, which provides a lower bound to the time required for a quantum state to evolve to an orthogonal state.

Under steady-state conditions, the classical dynamical activity $\mathcal{A}(\tau)$ increases linearly as a function of $\tau$, $\mathcal{A}(\tau) = \mathfrak{a}_\mathrm{ss}\tau$.
However, the quantum dynamical activity exhibits superlinear scaling within a certain interval, even under steady-state conditions, owing to coherent dynamics \cite{Nishiyama:2024:ExactQDAPRE}. 
In Ref.~\cite{Hasegawa:2020:QTURPRL}, an asymptotic expression of $\mathcal{B}(\tau)$ for $\tau \to \infty$ was derived, revealing that $\mathcal{B}(\tau)$ increases linearly as a function of $\tau$ for a sufficiently large $\tau$. 

\section{Derivation}

\subsection{Thermodynamic concentration inequality for quantum dynamics}
Here, we provide the proofs of Eqs.~\mainUresultUquantum{} and \mainUresultUclassical{}. 
From Eq.~\eqref{eq:Psi_tau_def},
the probability of no jump $\mathfrak{p}(\tau)$ is
\begin{align}
    \mathfrak{p}(\tau)=\braket{\psi_{S}(0)|\mathcal{V}_{\boldsymbol{0}}^{\dagger}\mathcal{V}_{\boldsymbol{0}}|\psi_{S}(0)},
    \label{eq:pfrak_def}
\end{align}
which satisfies the following inequality:
\begin{align}
    \left|\braket{\Psi(0)|\Psi(\tau)}\right|^{2}&=\left|\braket{\psi_{S}(0)|\mathcal{V}_{\boldsymbol{0}}|\psi_{S}(0)}\right|^{2}\nonumber\\&\le\left|\braket{\psi_{S}(0)|\mathcal{V}_{\boldsymbol{0}}^{\dagger}\mathcal{V}_{\boldsymbol{0}}|\psi_{S}(0)}\right|\nonumber\\&=\mathfrak{p}(\tau).
    \label{eq:fidelity_frakp}
\end{align}
For the first two lines, we used the Cauchy-Schwarz inequality. 
Moreover, from Eq.~\eqref{eq:QDA_Bures_angle}, we obtain the following relation for $0\le\frac{1}{2}\int_{0}^{\tau}\frac{\sqrt{\mathcal{B}(t)}}{t}dt\le\frac{\pi}{2}$:
\begin{align}
    \cos\left[\frac{1}{2}\int_{0}^{\tau}\frac{\sqrt{\mathcal{B}(t)}}{t}dt\right]\leq\left|\braket{\Psi(\tau)|\Psi(0)}\right|.
    \label{eq:fidelity_QDA}
\end{align}
From Eqs.~\eqref{eq:fidelity_frakp} and \eqref{eq:fidelity_QDA}, we obtain
\begin{align}
    \cos\left[\frac{1}{2}\int_{0}^{\tau}\frac{\sqrt{\mathcal{B}(t)}}{t}dt\right]^{2}\leq\mathfrak{p}(\tau).
    \label{eq:quantum_DACI}
\end{align}

Next, we associate $\mathfrak{p}(\tau)$ with probability $P(N(\tau)=0)$.
When there is no jump within the interval $[0,\tau]$, the observable is $N(\tau) = 0$ due to the condition in Eq.~\OvarnothingUdef{}; however, the inverse is not necessarily true. 
Subsequently, the following relation holds:
\begin{align}
    P(N(\tau) = 0) \ge \mathfrak{p}(\tau).
    \label{eq:P_mathfrakp_ineq}
\end{align}
When the weight vector $C_m$ is positive for all the elements, the equality in Eq.~\eqref{eq:P_mathfrakp_ineq} holds. 
Combining Eqs.~\eqref{eq:quantum_DACI} and \eqref{eq:P_mathfrakp_ineq} completes the proofs of Eqs.~\mainUresultUquantum{} and \mainUresultUclassical{}.

\subsection{Thermodynamic concentration inequality for classical dynamics}

Here, we provide the proof of Eq.~\mainUresultUclassicalUstronger{}. 
For a classical Markov process with a time-independent transition rate, 
the probability of no jump $\mathfrak{p}(\tau)$ is expressed as
\begin{align}
    \mathfrak{p}(\tau)=\sum_{\mu}P(\mu;0)e^{-\tau\sum_{\nu(\neq\mu)}W_{\nu\mu}}.
    \label{eq:mathfrakp_expr}
\end{align}
Applying the Jensen inequality, we obtain
\begin{align}
    \mathfrak{p}(\tau)&=\sum_{\mu}P(\mu;0)e^{-\tau\sum_{\nu(\neq\mu)}W_{\nu\mu}}\nonumber\\&\ge e^{-\tau\sum_{\mu}P(\mu;0)\sum_{\nu(\neq\mu)}W_{\nu\mu}}\nonumber\\&=e^{-\tau\mathfrak{a}(0)},
    \label{eq:mathfrakp_exp_ineq}
\end{align}
which, along with Eq.~\eqref{eq:P_mathfrakp_ineq}, completes Eq.~\mainUresultUclassicalUstronger{}. 

\subsection{Observable expectation bound}

Let $X$ be a random variable and $X_{\mathrm{max}}$ be its maximum value. 
Substituting these into Eq.~\MarkovUinequality{},
the following relation holds for $X_\mathrm{max} > a$:
\begin{align}
    P(X\le a)\leq\frac{\mathbb{E}[X_{\mathrm{max}}-X]}{X_{\mathrm{max}}-a}.
    \label{eq:reverse_Markov_ineq}
\end{align}
Equation~\eqref{eq:reverse_Markov_ineq} is  the reverse Markov inequality. 
Substituting $a = 0$, $X = |N(\tau)|$, and $X_{\mathrm{max}}=N_{\mathrm{max}}\equiv\max|N(\tau)|$ into Eq.~\eqref{eq:reverse_Markov_ineq},
we have
\begin{align}
    P(|N(\tau)|\le0)\leq1-\frac{\mathbb{E}[|N(\tau)|]}{N_{\mathrm{max}}}.
    \label{eq:reverse_Markov_ineq2}
\end{align}
Using $P(N(\tau)=0)=P(|N(\tau)|\le0)$ and Eqs.~\mainUresultUquantum{} and \mainUresultUclassicalUstronger{}, we obtain Eqs.~\thermoUMarkovUineq{} and \thermoUMarkovUineqUmathfraka{}.

\subsection{Correlation bound}

The bound for the correlation function was obtained in Ref.~\cite{Hasegawa:2024:TCI}. 
The concentration inequalities in Eq.~\thermoUMarkovUineqUmathfraka{} can also be used to derive the bound for the correlation function. 
We focus on the classical case [Eq.~\MasterUeqUdef{}].
We introduce the score function $S\left(B_\nu\right)$ that takes the state $B_\nu(\nu \in\{1,2, \ldots, M\})$ and outputs a real value within $(-\infty, \infty)$. We define $S_{\max } \equiv \max _{B \in \mathcal{B}}|S(B)|$ to
represent the highest absolute value of the score function within the set $\mathcal{B}$. Similarly, we introduce another score function $T\left(B_\nu\right)$ and define $T_{\max }$. 
Consider the correlation function $C(t)\equiv\mathbb{E}[S(X(0))T(X(t))]$, where
\begin{align}
    \mathbb{E}\left[S(X(0))T(X(t))\right]&=\sum_{\mu,\nu}T(B_{\nu})S(B_{\mu})P(\mu;0)P(\nu;t|\mu;0)\nonumber\\&=\mathds{1}\mathbf{T}e^{\mathbf{W}t}\mathbf{SP}(0).
    \label{eq:correlation_def}
\end{align}
Here, $\mathds{1} \equiv [1,1,\cdots,1]$ is the trace state and $\mathbf{S} \equiv \mathrm{diag}[S(B_1),\cdots,S(B_M)] $ 
and $\mathbf{T}\equiv \mathrm{diag}[T(B_1),\cdots,T(B_M)]$. 
For example, consider a classical system with two states $\mathcal{B}=$ $\left\{B_1, B_2\right\}$, where $X(t)$ exhibits random switching between $B_1$ and $B_2$.
Here, the score function is typically $S\left(B_1\right)=-1$ and $S\left(B_2\right)=-1$. 
The change in the correlation function given by $C(\tau)-C(0)$ can be expressed as $N(\tau)$, as follows:
\begin{align}
N(\tau)=S(X(0))T(X(\tau)) - S(X(0))T(X(0)).
    \label{eq:N_correlation_def}
\end{align}
$N(\tau)$ satisfies the condition in Eq.~\OvarnothingUdef{} and its expectation provides $\mathbb{E}[N(\tau)] = C(\tau)-C(0)$, which is the change in the correlation function. 
Substituting Eq.~\eqref{eq:N_correlation_def} into Eq.~\thermoUMarkovUineqUmathfraka{}, we derive the upper bound for the change in the correlation function as follows:
\begin{align}
    \left|C(0)-C(\tau)\right|\le2S_{\mathrm{max}}T_{\mathrm{max}}\left(1-e^{-\mathfrak{a}(0)\tau}\right).
    \label{eq:correlation_bound_at}
\end{align}

\subsection{Linear response bound}

Here, we derive the bound for a weak perturbation applied to the system, based on Ref.~\cite{Hasegawa:2024:TCI}. 
The steady-state probability distribution satisfies
\begin{align}
    \mathbf{W}\mathbf{P}_\mathrm{ss} = 0,
    \label{eq:steadystate_condition}
\end{align}
where $\mathbf{P}_{\mathrm{ss}} \equiv [P_\mathrm{ss}(1),\cdots,P_\mathrm{ss}(M)]^\top$. 
Suppose a weak perturbation is applied to the master equation of Eq.~\MasterUeqUdef{}, which corresponds to the following replacement in Eq.~\MasterUeqUdef{}:
\begin{align}
    \mathbf{W}\to\mathbf{W}+\chi\mathbf{F}h(t),
    \label{eq:perturbation_W}
\end{align}
where $\chi > 0$ is the perturbation strength, $\mathbf{F}$ is an $M\times M$ real matrix, and $h(t)$ is a real function of $t$. 
Assuming that $\chi$ is sufficiently weak, we employ the perturbation expansion:
\begin{align}
    \mathbf{P}(t)=\mathbf{P}_{\mathrm{ss}}+\chi\mathbf{P}_{1}(t)+O(\chi^{2}),
    \label{eq:Pt_expand}
\end{align}
where $\mathbf{P}_1(t)$ is a first-order term. 
Substituting Eqs.~\eqref{eq:perturbation_W} and \eqref{eq:Pt_expand} into Eq.~\MasterUeqUdef{} and retaining the terms up to the first order in $\chi$, we obtain the following differential equation with respect to $\mathbf{P}_1(t)$:
\begin{align}
    \frac{d}{dt}\mathbf{P}_{1}(t)=\mathbf{W}\mathbf{P}_{1}(t)+\mathbf{F}\mathbf{P}_{\mathrm{ss}}h(t).
    \label{eq:P1_ODE}
\end{align}
Solving Eq.~\eqref{eq:P1_ODE}, the solution is 
\begin{align}
    \mathbf{P}_{1}(t)=\int_{-\infty}^{t}e^{\mathbf{W}\left(t-t^{\prime}\right)}\mathbf{F}\mathbf{P}_{\text{ss }}h(t^{\prime})dt^{\prime}.
    \label{eq:P1_solution}
\end{align}
Consider the expectation of the score function $G(B)$ for $B \in \mathfrak{B}$, and consider its deviation from the steady-state condition under perturbation:
\begin{align}
   \Delta G(t)&\equiv\sum_{\mu}P(\mu;t)G(B_{\mu})-\sum_{\mu}P_{\mathrm{ss}}(\mu)G(B_{\mu})\nonumber\\&=\chi\int_{-\infty}^{\infty}R_{G}(t-t^{\prime})h(t^{\prime})dt^{\prime},
    \label{eq:DeltaT_def}
\end{align}
where $R_G(t)$ is the linear response function given by
\begin{align}
    R_G(t)=\begin{cases}
\mathds{1}\mathbf{G}e^{\mathbf{W}t}\mathbf{F}\mathbf{P}_{\mathrm{ss}} & t\geq0,\\
0 & t<0.
\end{cases}
\label{eq:linear_response_func_def}
\end{align}
Knowing the linear response function $R_G(t)$, we can evaluate any input-output relation in the linear response regime. 
Next, we associate the linear response function with the correlation function. 
Using matrix notation, the correlation function can be expressed by \cite{Hasegawa:2024:TCI}
\begin{align}
    C(t)=\mathds{1}\mathbf{T}e^{\mathbf{W}t}\mathbf{S}\mathbf{P}_{\mathrm{ss}}.
    \label{eq:Ct_matrix_expr}
\end{align}
From Eq.~\eqref{eq:Ct_matrix_expr}, the time derivative of $C(t)$ is
\begin{align}
    \frac{d}{dt}C(t)=\mathds{1}\mathbf{T}e^{\mathbf{W}t}\mathbf{W}\mathbf{S}\mathbf{P}_{\mathrm{ss}}.
    \label{eq:Ct_time_derivative}
\end{align}
Comparing Eq.~\eqref{eq:linear_response_func_def} with Eq.~\eqref{eq:Ct_time_derivative}, when the system satisfies $\mathbf{G}=\mathbf{T}$ and $\mathbf{F}=\mathbf{W}\mathbf{S}$, then the linear response function $R_T(t)$ is given by $\partial_t C(t)$. 
Suppose a constant perturbation is initiated at $t=0$.
This perturbation can be modeled as $h(t)=\Theta(t)$,
where $\Theta(t)$ denotes the Heaviside step function. 
Therefore, $\Delta T$ can be expressed as
\begin{align}
    \Delta T(t)&=\chi\int_{-\infty}^{\infty}R_{T}\left(t-t^{\prime}\right)\Theta\left(t^{\prime}\right)dt^{\prime}\nonumber\\&=\chi\int_{0}^{t}R_{T}\left(t-t^{\prime}\right)dt^{\prime}\nonumber\\&=\chi(C(t)-C(0)).
    \label{eq:DeltaT_by_Ct}
\end{align}
Using Eqs.~\eqref{eq:correlation_bound_at} and \eqref{eq:DeltaT_by_Ct}, we obtain
\begin{align}
    \left|\Delta T(\tau)\right|\leq2\chi S_{\max}T_{\max}\left(1-e^{-\mathfrak{a}_{\mathrm{ss}}\tau}\right),
    \label{eq:linear_response_bound}
\end{align}
where $\mathfrak{a}_\mathrm{ss}$ is the dynamical activity under steady-state conditions. 
Equation~\eqref{eq:linear_response_bound} is tighter than the bound presented in Ref.~\cite{Hasegawa:2024:TCI}.

\section{Comparison of bounds\label{app:comparison}}

In Eqs.~\mainUresultUclassical{} and \mainUresultUclassicalUstronger{}, we show two bounds that hold for classical Markov processes. 
Generally, it is not clear whether Eq.~\mainUresultUclassical{} or Eq.~\mainUresultUclassicalUstronger{} provides a tighter bound. However, under the steady-state condition, Eq.~\mainUresultUclassicalUstronger{} is tighter, which is shown below. 

The inequality that needs to be proved is given by
\begin{align}
    \cos\left[\frac{1}{2}\int_{0}^{\tau}\frac{\sqrt{\mathcal{A}(t)}}{t}dt\right]^{2}\le e^{-\mathfrak{a}(0)\tau}.
    \label{eq:cos_a0_ineq}
\end{align}
In the steady-state scenario, the dynamical activity is $\mathcal{A}(\tau) = \mathfrak{a}_\mathrm{ss}\tau$.
This indicates that the inequality to be proved can be written as
\begin{align}
    \cos(a)\le\exp\left[-\frac{a^{2}}{2}\right],
    \label{eq:inequality_of_interest}
\end{align}
where $0 \le a \le \pi/2$. Let us define
\begin{align}
    R(a)\equiv\ln\cos(a)+\frac{a^{2}}{2}.
    \label{eq:Rs_def}
\end{align}
Because its derivative is $R'(a) = a - \tan(a) \le 0$ for $0 \le a \le \pi/2$, 
we prove Eq.~\eqref{eq:cos_a0_ineq} under the steady-state condition.

\section{Numerical simulation}

We perform quantum and classical numerical simulations to validate the derived uncertainty relations. 
Specifically, we focus on the inequalities relating the no-jump probability and the dynamical activities [Eqs.~\eqref{eq:quantum_DACI} and \eqref{eq:mathfrakp_exp_ineq}], and those relating the $p$-norm and the dynamical activities [Eqs.~\pUnormUTUR{} and \pUnormUTURUstrongerUclassical{}]. 
Note that validating Eqs.~\eqref{eq:quantum_DACI} and \eqref{eq:mathfrakp_exp_ineq} directly leads to the validation of Eqs.~\mainUresultUquantum{} and \mainUresultUclassicalUstronger{}.
Furthermore, as indicated in the main text, when $p \to \infty$, Eqs.~\pUnormUTUR{} and \pUnormUTURUstrongerUclassical{} reduce to Eqs.~\thermoUMarkovUineq{} and \thermoUMarkovUineqUmathfraka{}, respectively. 
Hence, confirming Eqs.~\pUnormUTUR{} and \pUnormUTURUstrongerUclassical{} for large values of $p$ inherently validates Eqs.~\thermoUMarkovUineq{} and \thermoUMarkovUineqUmathfraka{}.

\subsection{Quantum case}

For the quantum model simulation, we use a two-level atom driven by a classical laser field, as described by the Lindblad equation. The system Hamiltonian and the jump operator are given by
\begin{align}
H_{S}&=\Delta\ket{\mathfrak{e}}\bra{\mathfrak{e}}+\frac{\Omega}{2}(\ket{\mathfrak{e}}\bra{\mathfrak{g}}+\ket{\mathfrak{g}}\bra{\mathfrak{e}}),\label{eq:HS_def}\\L&=\sqrt{\kappa}\ket{\mathfrak{g}}\bra{\mathfrak{e}}.
\label{eq:L_def}
\end{align}
Here, $\ket{\mathfrak{e}}$ and $\ket{\mathfrak{g}}$ represent the excited and ground states, respectively. The parameters $\Delta$, $\Omega$, and $\kappa$ correspond to the detuning between the laser and atomic frequencies, the Rabi oscillation frequency, and the decay rate, respectively. The jump operator $L$ induces the transitions from the excited state $\ket{\mathfrak{e}}$ to the ground state $\ket{\mathfrak{g}}$.

We first validate the inequality given by Eq.~\eqref{eq:quantum_DACI}, which leads to the first main result of Eq.~\mainUresultUquantum{} in the main text. 
We randomly choose the model parameters ($\Delta$, $\Omega$, $\kappa$, and $\tau$) and the initial density operator $\rho_S(0)$, and then calculate $\mathfrak{p}(\tau)$ and $\cos[(1/2)\int_{0}^{\tau}\sqrt{\mathcal{B}(t)}/t\,dt]^{2}$ for each configuration. 
In Fig.~\ref{fig:quantum_TUR}(a), we plot $\mathfrak{p}(\tau)$ against $\cos[(1/2)\int_{0}^{\tau}\sqrt{\mathcal{B}(t)}/t\,dt]^{2}$ with the points, where the dashed line denotes the equality case of Eq.~\eqref{eq:quantum_DACI}. 
The quantum dynamical activity $\mathcal{B}(\tau)$ is evaluated using the expression of Eq.~\eqref{eq:QDA_exact_solution}. 
Since all points lie above the dashed line, 
Fig.~\ref{fig:quantum_TUR}(a) provides numerical validation for the inequality \eqref{eq:quantum_DACI}.
In regions where $\mathfrak{p}(\tau)$ is large, it can be seen that there are many points near the dashed line, indicating that the inequality is quite tight. On the other hand, in regions where $\mathfrak{p}(\tau)$ is small, it is not as tight.

Next, we validate Eq.~\pUnormUTUR{}.
Again, we randomly choose the model parameters ($\Delta$, $\Omega$, $\kappa$, and $\tau$) and the initial density operator $\rho_S(0)$.
Then, we calculate the $p$-norm $\|N(\tau)\|_{p}/\|N(\tau)\|_{1}$ by numerically solving the stochastic Schr\"odinger equation corresponding to the Lindblad equation \superopUdef{} within the interval $[0,\tau]$. 
Here, we employ the counting observable defined by Eq.~\MUdef{} for $N(\tau)$. 
In Figs.~\ref{fig:quantum_TUR}(b)--(d), 
$\|N(\tau)\|_{p}/\|N(\tau)\|_{1}$ is plotted against $\sin[(1/2)\int_{0}^{\tau}\sqrt{\mathcal{B}(t)}/t\,dt]^{-2(p-1)/p}$ with points for three different $p$ values: (b) $p=2$, (c) $p=4$, and (d) $p=8$. 
The $p=2$ case directly leads to the bound for the scaled variance $\mathrm{Var}[N(\tau)]/\mathbb{E}[N(\tau)]^2$, which is the quantity considered in the thermodynamic uncertainty relation. 
In Figs.~\ref{fig:quantum_TUR}(b)--(d), 
the dashed line shows the equality case of Eq.~\pUnormUTUR{}.
Because all points are above the dashed line, 
Figs.~\ref{fig:quantum_TUR}(b)--(d) numerically confirm that Eq.~\pUnormUTUR{} holds for $p=2,4$, and $8$.

\subsection{Classical case}

For the simulation of the classical model, we consider a Markov process with $M$ states, governed by Eq.~\MasterUeqUdef{} in the main text.

\begin{figure}
    \includegraphics[width=1\linewidth]{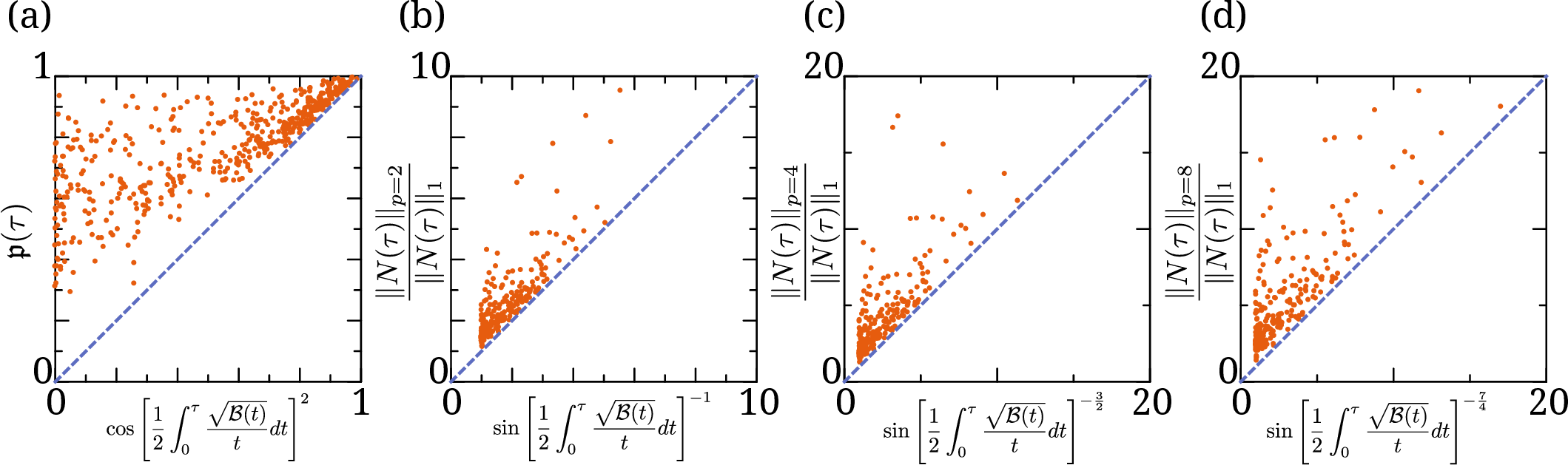}
    \caption{
Results of numerical simulation in the quantum system. 
(a)
Verification of Eq.~\eqref{eq:quantum_DACI}. 
$\mathfrak{p}(\tau)$ [Eq.~\eqref{eq:pfrak_def}], which is the probability that there is no jump within $[0,\tau]$, is plotted against $\cos[(1/2)\int_{0}^{\tau}\sqrt{\mathcal{B}(t)}/tdt]^{2}$. 
The points denote random realizations and the dashed line is the equality case of Eq.~\eqref{eq:quantum_DACI}. 
(b)--(d) 
Verification of Eq.~\pUnormUTUR{} for different $p$ values: (b) $p=2$, (c) $p=4$, and (d) $p=8$. 
The $p$-norm $\|N(\tau)\|_{p}/\|N(\tau)\|_{1}$ is plotted against (b) $\sin[(1/2)\int_{0}^{\tau}\sqrt{\mathcal{B}(t)}/t\,dt]^{-1}$, (c) $\sin[(1/2)\int_{0}^{\tau}\sqrt{\mathcal{B}(t)}/t\,dt]^{-3/2}$, and (d) $\sin[(1/2)\int_{0}^{\tau}\sqrt{\mathcal{B}(t)}/t\,dt]^{-7/4}$. 
Again, random realizations are shown with points and the solid line indicates the equality case of Eq.~\pUnormUTUR{}. 
In (a)--(d), quantum simulations are performed by choosing values for $\Delta$, $\Omega$, and $\kappa$ between $0.1$ and $3$, and setting $\tau$ between $0.1$ and $\sqrt{10}$.
Note that points with $(1/2)\int_{0}^{\tau}\sqrt{\mathcal{B}(t)}/t\,dt > \pi/2$ are not shown. 
$\|N(\tau)\|_p$ is estimated by averaging the results of $10^5$ simulations.
    }
    \label{fig:quantum_TUR}
\end{figure}

\begin{figure}
    \includegraphics[width=1\linewidth]{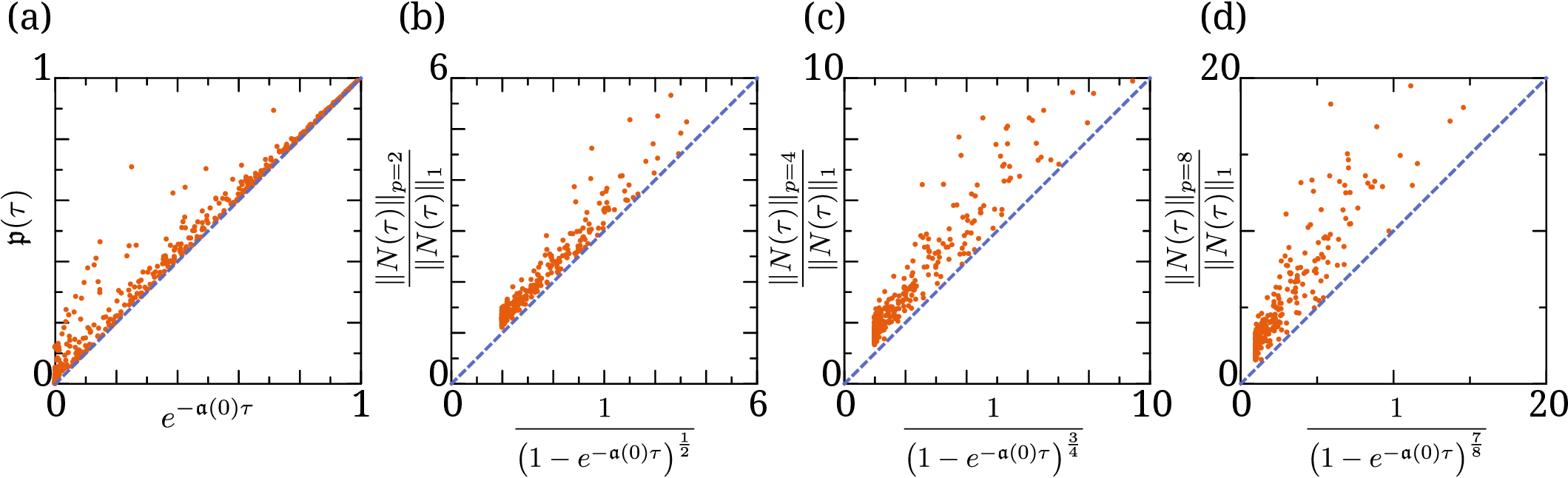}
    \caption{
Results of numerical simulation in the classical system. 
(a)
Verification of Eq.~\eqref{eq:mathfrakp_exp_ineq}. 
$\mathfrak{p}(\tau)$ [Eq.~\eqref{eq:mathfrakp_expr}] is plotted against $e^{-\mathfrak{a}(0) \tau}$. 
The points denote random realizations and the dashed line is the equality case of Eq.~\eqref{eq:mathfrakp_exp_ineq}. 
(b)--(d) 
Verification of Eq.~\pUnormUTURUstrongerUclassical{} for different $p$ values: (b) $p=2$, (c) $p=4$, and (d) $p=8$. 
The $p$-norm $\|N(\tau)\|_{p}/\|N(\tau)\|_{1}$ is plotted against (b) $(1-e^{-\mathfrak{a}(0) \tau})^{-1/2}$, (c) $(1-e^{-\mathfrak{a}(0) \tau})^{-3/4}$, and (d) $(1-e^{-\mathfrak{a}(0) \tau})^{-7/8}$. 
Again, random realizations are shown with points and the solid line indicates the equality case of Eq.~\pUnormUTURUstrongerUclassical{}. 
In (a)--(d), classical simulations begin by selecting $M$ values between $2$ and $5$. Then, the transition rate $\mathbf{W}$ is randomly produced, and $\tau$ is chosen within the range of $0.1$ to $10$.
$\|N(\tau)\|_p$ is estimated by averaging the results of $10^6$ simulations.
    }
    \label{fig:classical_TUR}
\end{figure}

We first validate the inequality given by Eq.~\eqref{eq:mathfrakp_exp_ineq}, which leads to Eq.~\mainUresultUclassicalUstronger{}. 
We select $M$ at random, and generate a random transition matrix $\mathbf{W}$ and a random initial probability $\mathbf{P}(0)$. Subsequently, we calculate $\mathfrak{p}(\tau)$ and $e^{-\mathfrak{a}(0)\tau}$ for every configuration.
In Fig.~\ref{fig:classical_TUR}(a), we plot $\mathfrak{p}(\tau)$ against $e^{-\mathfrak{a}(0) \tau}$ with the points, where the dashed line corresponds to the equality case of Eq.~\eqref{eq:mathfrakp_exp_ineq}. 
Because all points lie above the dashed line, Fig.~\ref{fig:classical_TUR}(a) offers numerical evidence supporting the inequality \eqref{eq:mathfrakp_exp_ineq}.
Compared with Fig.~\ref{fig:quantum_TUR}(a), it is clear that the inequality is tighter. This is particularly noticeable in regions where $\mathfrak{p}(\tau)$ is small.

Next, we validate Eq.~\pUnormUTURUstrongerUclassical{}.
Again, we select $M$ at random, and generate a random transition matrix $\mathbf{W}$ and a random initial probability $\mathbf{P}(0)$. 
Then, we calculate the $p$-norm $\|N(\tau)\|_{p}/\|N(\tau)\|_{1}$ by numerically solving the Gillespie algorithm for the master equation \MasterUeqUdef{} within the interval $[0,\tau]$. 
Here, we employ the counting observable defined by Eq.~\MUdef{} for $N(\tau)$, where the weight vector $C_m$ is determined randomly. 
In Figs.~\ref{fig:classical_TUR}(b)--(d), 
$\|N(\tau)\|_{p}/\|N(\tau)\|_{1}$ is plotted against $(1-e^{-\mathfrak{a}(0)\tau})^{-(p-1)/p}$ with points for three different $p$ values: (b) $p=2$, (c) $p=4$, and (d) $p=8$. 
In Figs.~\ref{fig:classical_TUR}(b)--(d), 
the dashed line shows the equality case of Eq.~\pUnormUTURUstrongerUclassical{}.
Because all points are above the dashed line, 
Figues~\ref{fig:classical_TUR}(b)--(d) numerically confirm that Eq.~\pUnormUTURUstrongerUclassical{} holds for $p=2,4$, and $8$.

